\documentclass[structabstract]{aa}
\usepackage{natbib}
\usepackage{graphicx}
\usepackage{amssymb}
\usepackage{amsmath}

\usepackage{txfonts}

\begin{document}
   \title{Acceleration of cosmic rays by young \\
core-collapse supernova remnants.}

   \subtitle{}

   \author{I.Telezhinsky
          \inst{1,2}\fnmsep\thanks{\email{igor.telezhinsky@desy.de}}
          \and
	  V.V. Dwarkadas
          \inst{3}
	  \and
          M. Pohl
          \inst{1,2}
          }

   \institute{DESY, Platanenallee 6, 15738 Zeuthen, Germany
         \and
            University of Potsdam, Institute of Physics \& Astronomy, Karl-Liebknecht-Strasse 24/25, 14476 Potsdam, Germany 
         \and            
             University of Chicago, Department of Astronomy\&Astrophysics, 5640 S Ellis Ave, AAC 010c, Chicago, IL 60637, U.S.A. 
             }

   \date{Received ; accepted }


  \abstract
{Supernova remnants (SNRs) are thought to be the primary candidates
for the sources of Galactic cosmic rays. According to the
diffusive shock acceleration theory,  SNR shocks produce a power-law
spectrum with an index of s = 2, perhaps nonlinearly modified to 
harder spectra at high energy. Observations of SNRs often indicate
particle spectra that are softer than that and show features not expected
from classical theory. Known drawbacks of the standard approach are
the assumption that SNRs evolve in a uniform environment, and that
the reverse shock does not accelerate particles. Relaxing these
assumptions increases the complexity of the problem, because one
needs reliable hydrodynamical data for the plasma flow as well as
good estimates for the magnetic field (MF) at the reverse shock.}  {We
show that these two factors are especially important when modeling
young core-collapse SNRs that evolve in a complicated circumstellar
medium shaped by the winds of progenitor stars.} {We used
high-resolution numerical simulations for the hydrodynamical
evolution of the SNR. Instead of parametrizations of
the MF profiles inside the SNR, we followed the
advection of the frozen-in MF inside the
SNR, and thus obtained the B-field value at all locations, in particular at the
reverse shock. To model cosmic-ray acceleration we solved the cosmic-ray
transport equation in test-particle approximation.} {We find that
the complex plasma-flow profiles of core-collapse SNRs significantly
modify the particle spectra. Additionally, the reverse shock
strongly affects the emission spectra and the surface brightness.}
   {}

\authorrunning{I. Telezhinsky et al.}
\titlerunning{Acceleration of CRs by young core-collapse SNRs.}

\maketitle
%

\section{Introduction}

It is often assumed that cosmic rays (CRs) with energies below the
knee of the CR spectrum ($\sim 10^{16}$~eV) are accelerated in
supernova remnants (SNRs). If this were correct, particle spectra
should follow a power-law with index, $s=2$, with an exponential
cut-off at very high energy. However, the wealth of recent
observational data on, e.g., RX~J0852.0-4622 \citep{Ahaetal07b},
RCW~86 \citep{Ahaetal09}, SN~1006 \citep{Aceetal2010_1006}, Cas~A
\citep{Accetal10_CasA, Abdetal10_CasA}, and Tycho's SNR
\citep{Accetal11_Tycho, Gioetal11} provide strong evidence that the
particle spectra are significantly softer.  Moreover, the spectral
shape is not a pure power law with exponential cut-off, as for example
in Cas~A. In addition, multi-wavelength images of SNRs puzzle
observers and force them to introduce multi-zone emission models
\citep{AraCui10, Lemetal12, AtoDer12}.

Several groups have introduced modifications to classical diffusive
shock acceleration (DSA) theory \citep{Axfetal77, Kry77, Bel78,
  BlaOst78} and its nonlinear counterpart (NDSA) \citep{MalDru01}.
\citet{Maletal12} claim that ion-neutral collisions in the SNR
vicinity may steepen the energy spectrum of particles by one power
with respect to classical DSA due to an evanescence of Alfv\'en waves
that permits the escape of particles in a certain momentum range.
\citet{Blaetal12} discuss the spectral softening that arises from the
energy and momentum transfer of neutrals that return to the upstream
region.  The mean free path of neutrals is longer than the
length-scale of shock modification because of CR streaming, and
hence the effective compression ratio is $< 4$ and the particle
spectrum is soft even for high-energy particles. \citet{Cap12} argues
that in the upstream region the scattering centers propagate with Alfv\'en velocity,
which in the presence of very strong MF amplification
reduces the compression ratio and leads to soft spectra of particles
\citep[see also][]{2000JPlPh..64..459L}. According to this study, the
acceleration efficiency saturates at around 30\%. \citet{Inoetal10}
consider multiple weak secondary shocks that appear when the primary
forward shock propagates through a medium filled with small dense
cloudlets. The CRs are re-accelerated at these low-Mach-number shocks,
resulting in a softer spectrum. The largest and strongest secondary
shock in the system is the reverse shock (RS) that propagates through
the ejecta.  High-resolution radio and X-ray observations support the
notion that particle acceleration can also occur at RS
\citep{Gotetal01,Rhoetal02,Deletal02,Sasetal06,HelVin08}, but only
recently has this possibility been considered in theoretical
calculations \citep{ZirAha10,ZirPtu11, Teletal12a, Teletal12b}, which
relax the restricting assumptions about the magnetic field (MF) in the RS
region \citep[e.g.][]{Elletal05}. The RS accelerates particles of the
ejecta and thus provides a second population of relativistic particles
in addition to that produced at the forward shock. The superposition of these
two components may lead to spectral modifications in the
volume-integrated emission from SNRs that are not easily reproduced
with so-called one-zone/one-shock models, and accordingly, multiple
zones/particle populations need to be employed \citep{AraCui10,
  Lemetal12, AtoDer12}. Misinterpretations are possible
if the forward shock (FS) {is considered} the sole accelerator.

Another idealization is the assumption that SNRs evolve in a uniform
environment. Given that stellar mass-loss considerably modifies
  the surroundings of stars, as well as known inhomogeneities in the
  interstellar medium, it is clear that this cannot hold, and
calculations have started to take into account the complicated nature
of the SNR environment \citep{EllByk11, ZirPtu11, Elletal12,
  Teletal12b}. This increases the complexity of the problem, because
one requires reliable hydrodynamics of the plasma flow as well as
knowledge about the MF at the reverse shock.

In this paper we show that taking into account both points, namely the
acceleration of particles at the RS and the complex hydrodynamics of
SNRs, is particularly important for modeling young core-collapse SNRs
that evolve in a circumstellar medium shaped by the wind from the
progenitor star. The resulting volume-integrated spectrum of particles
and their consequent radiation is significantly different from that of
planar-shock calculations and dependent on the type of SNR. To
accomplish our study, we perform high-resolution simulations of the
hydrodynamical evolution of the SNR with initial and environmental
conditions representative of Type Ic and Type IIP supernovae (SNe). We
consider the transport of frozen-in MF by the plasma flow inside the
SNR to trace its evolution inside the remnant and particularly in the
RS region separately for the radial and the tangential
field. We model cosmic-ray acceleration by solving the cosmic-ray
transport equation in test-particle approximation. Finally, we
calculate the resulting emission from the SNR and construct
surface-brightness maps in various energy bands. Thus we trace the
complex particle distribution resulting from particle acceleration at
both shocks.

\section{Hydrodynamics}
\label{sec:hd}

Our goal is to study how the evolution of SNR shock waves into
  the medium created by their progenitor stars modifies the spectra
  and high-energy emission from these objects. Given the impossibility
  of exploring the total diversity of paths that can lead to a
  supernova explosion, we have instead chosen to sample two
  distinctive types of SNe.  These illustrate the detailed structure
  of the circumstellar medium into which SNe advance, and also how the
  various features in the spectra, and the very high energy emission,
  relate to the evolutionary properties of the SN.

We investigate here Type Ic and Type IIP SNe. We chose these
because they show two different regimes of evolution, and because many
SNe, even those of other types, will fall somewhere close to or in
between these types. For instance, Cas A is classified as a Type IIb SN
based on spectra obtained from light echoes. However, its evolution in
a red supergiant (RSG) wind \citep{CheOis03}, within which it is probably
currently expanding, would be quite similar to the evolution in a wind
medium for a Type IIP SN. Thus we hope to give an overview of the spectra
and evolution of many different types of young SNRs via these
calculations. Below we describe the assumptions that went into the
simulations, and the evolution of the remnant in each case.

To simulate the SN expansion into the ambient medium, we
first need to understand the nature of the ambient medium. Since the
medium around core-collapse SNe is shaped mainly by mass-loss from the
progenitor star, we need to take into account the evolution of
the progenitors of these SNe.

\begin{figure}[!t]
\vspace{5mm}
\centering
\includegraphics[width=0.48\textwidth]{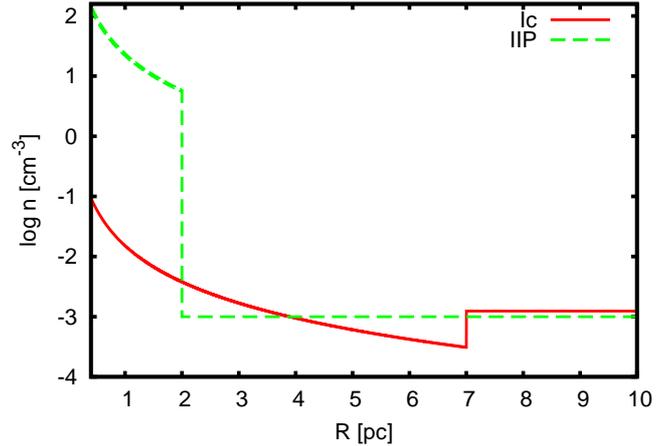}
\caption{Density profiles of circumstellar gas before the explosion of Type-Ic
  and Type-IIP SNe.}
\label{csm}
\end{figure}

Type IIP SNe arise from the explosion of RSGs. In general, these
  stars start as O- or B-type stars while on the main sequence, with
  initial masses between about 8 and 30 solar masses. In the
main-sequence stage these stars lose mass via radiatively driven
winds, with high wind velocities ($ > 1000\ {\rm km\ s}^{-1}$) with a
mass-loss rate of about 10$^{-7}$ M$_{\odot}\ {\rm yr}^{-1}$. The
interaction of the main-sequence wind with the surrounding
(constant-density) interstellar medium gives rise to a wind-blown
bubble \citep{WeaMcc77, SegLan96, MarLan05, Dwa05, Dwa07, ToaArt11,
  DwaRos12}. The standard structure of the wind bubble, going
outwards in radius from the star, consists of a freely expanding wind that
ends in a wind termination shock, a hot, low-density shocked wind
medium, a contact discontinuity, outer shock, and the external
medium. If the wind has constant parameters, the freely expanding wind
will have a density profile that decreases as r$^{-2}$, while the
shocked wind will have a more or less constant density. It is possible
that, depending on the surface temperature and number of ionizing
photons emitted from the star, there is a dense ionized (HII) region
inside the constant discontinuity \citep[see e.g.][]{Dwa11}.
Since this does not happen in all cases, we have not
taken it into account.

As the star moves off the main sequence into the RSG stage, it grows
considerably in size, the wind mass-loss rate increases to about
5$\times 10^{-5} M_{\odot} {\rm yr}^{-1}$ while the velocity drops to
a low value of about 10 km s$^{-1}$. This results in a new pressure
equilibrium. The high density ($ \propto
\dot{M}/v_w$) of the RSG wind leads to the formation of a wind region
with density almost four orders of magnitude above that of the
main-sequence wind. The RSG wind can also end in a shell, but we
chose for simplicity to exclude this. This leads to the density
structure shown in Fig~\ref{csm}, with an initial high-density wind
with density decreasing as r$^{-2}$, followed by a steep drop in
density toward the main-sequence wind zone, ending in the
main-sequence shell.

\begin{figure*}[!t]
\vspace{5mm}
\centering
\includegraphics[width=0.98\textwidth]{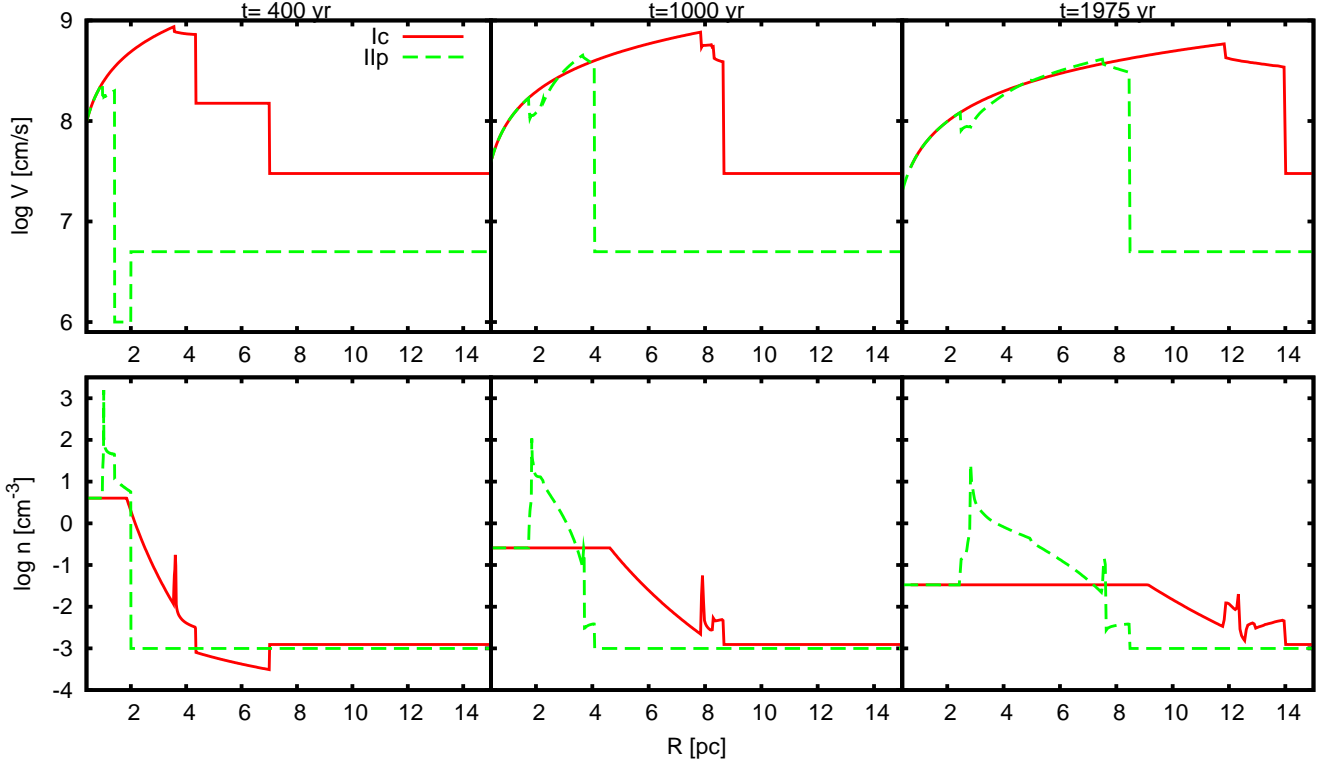}
\caption{Time evolution of the plasma outflow velocity (top) and density
(bottom) as a function of radius for Type-Ic and Type-IIP SNRs.}
\label{hd}
\end{figure*}

If the initial mass of the star exceeds about 30 M$_{\odot}$
  (depending on which models are used), the star will not end its
  life as a RSG.  It will leave the main sequence to become an RSG, or,
  in stars with initial mass $>50\ M_\odot$, to
  become a luminous blue variable. Following this, it may shed the
  outer H and perhaps He envelope, turning into a Wolf-Rayet (WR)
  star. The mass-loss rates of WR stars are somewhat lower than
those of RSG winds, but their wind velocities are more than two orders
of magnitude higher, leading to wind densities that are more than two
orders of magnitude lower. The high momentum of the winds pushes
outward on the RSG shell, breaking it up in the process and mixing
the RSG material into the WR wind \citep{Dwa07, ToaArt11,
  DwaRos12}. This mixed material approaches, and could bounce back
from, the main-sequence shell. Eventually, the system reaches an
equilibrium situation that in many ways resembles the main-sequence
bubble (and has almost the same radius, since the main sequence-shell
is very dense and expanding very slowly, on the order of 20-50 km s$^{-1}$),
but the extent of the freely expanding wind and shocked wind, and the
radius of the wind termination shock, are different. This leads to the
situation shown in Fig~\ref{csm}.

To explore the situation without having to simulate the
  entire evolution of the progenitor star and its wind medium, we
 approximated the wind density profile using reasonable average
  parameters.  Following this, we investigated the evolution of the SN
  shock wave within the medium. We used the VH-1 code, a 3D
finite-difference hydrodynamic code based on the piecewise parabolic
method \citep{ColWoo84}. Since we aim to study the effects of the
complicated environment, we assumed similar characteristics for
the SN density structure in each case, so that the differences in the
  evolution result purely from the difference in the ambient medium.
In each case we assumed an ejecta mass of about 5 M$_{\odot}$ and an
explosion energy of 10$^{51}$ ergs. The ejecta density is flat where
the plasma flow velocity is below a certain value, $u_{fl}$, and decreases as a
power law with radius, $\rho_{ej} \propto r^{-9}$, where the flow velocity is 
above $u_{fl}$ \citep{CheFra94, Dwa05}. The interaction of
the ejecta with the wind medium sets up a double-shock structure as
expected, consisting of a forward and reverse shock separated by a
contact discontinuity. The region between the outer shock and
  contact discontinuity contains shocked surrounding medium, whereas
  the region between the inner shock and contact discontinuity includes shocked
  ejecta. The structure of this shocked region depends on the ejecta
  density and the density profile of the surrounding medium
  \citep{Che82,Dwa11}, and has implications for the emission from the
  remnant.

The shock expansion in the wind medium in both cases is quite
similar initially, except that the shock moves slower in the
high-density region, as expected. The evolution changes once the shock
reaches the end of the freely expanding wind region. In the RSG case
one finds a huge drop in density beyond the wind region, whereas
in the WR case one finds an increase in density by a factor of
4. This change in density and the transition from the wind zone to a
medium of almost constant density destroys the self-similarity of the
solution. The interaction of the SNR forward shock with the wind
termination shock leads to a reflected shock that travels back into the
ejecta in the case of the Type Ic SN. In the case of the Type IIP, the
steep drop leads to the formation of a complicated ejecta
structure. These structures are used to compute the acceleration of
particles at the shock fronts. The evolution of the plasma velocity
and density profiles is shown in Fig.~\ref{hd}.

\begin{figure*}[!t]
\vspace{5mm}
\centering
\includegraphics[width=0.98\textwidth]{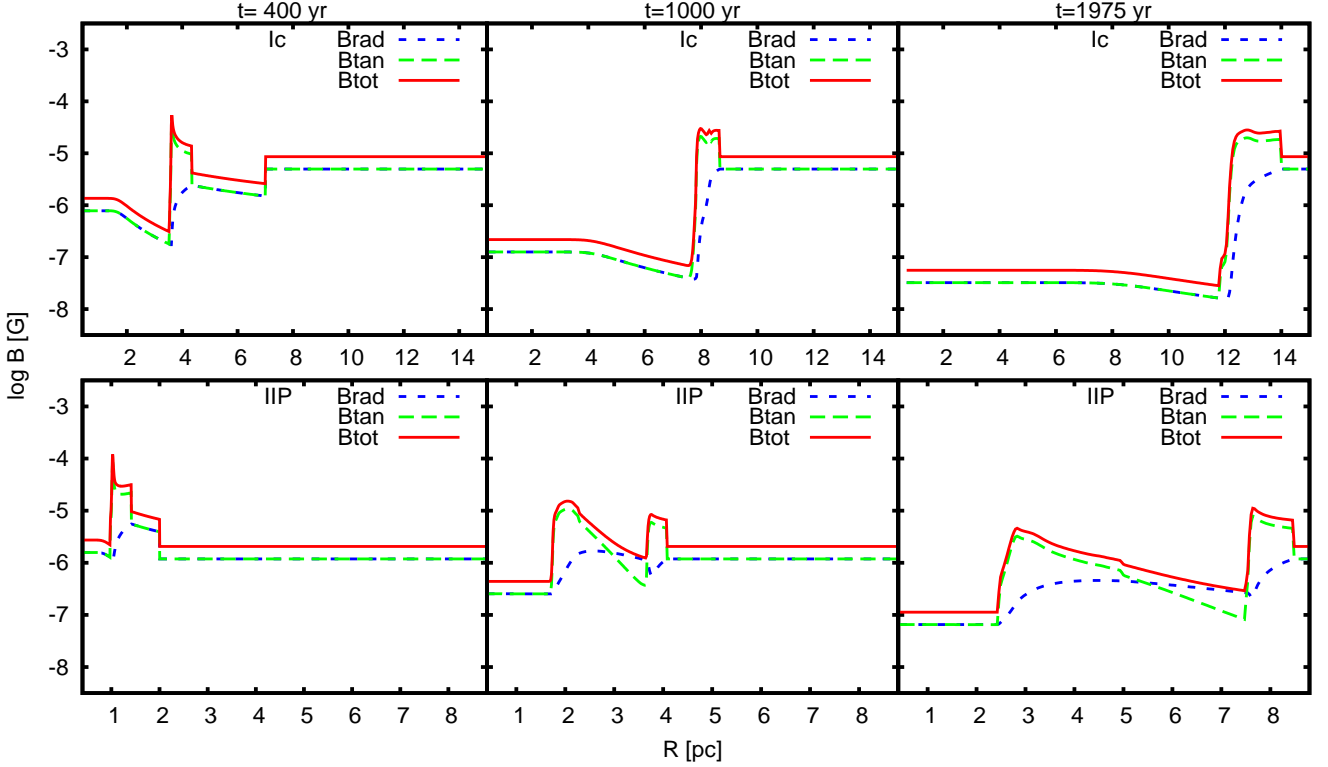}
\caption{Time evolution of the total, tangential, and radial MF as a function of radius
for Type-Ic (top) and Type-IIP (bottom) SNRs.}
\label{mf}
\end{figure*}

\section{Magnetic field}
\label{sec:mf}

The acceleration and subsequent radiation of relativistic particles significantly
depends on the distribution of MF in the SNR, especially
in the upstream and downstream vicinities of the shocks. In our
previous work \citep{Teletal12a, Teletal12b}, we assumed 
parametrizations for the MF inside the remnant following either
pressure or density distributions. We found that these simple
parametrizations fail for Type-IIP SNRs, since the
energy contained in MF would be roughly equivalent to the explosion
energy after a few hundreds years of evolution. In this work we implement a more realistic description of the MF distribution inside SNRs,
which is especially important for the acceleration of particles at the
RS, a key ingredient of our studies.

We start with the assumption that the SNR is filled with a perfectly
conducting fluid, which is common in astrophysical plasmas. The
evolution of the frozen-in MF satisfies the equation

\begin{equation}
\frac{\partial \mathbf{B}}{\partial t} = \bigtriangledown \times (\mathbf{u} \times \mathbf{B}),
\label{vecprod}
\end{equation}
where $\mathbf{B}$ is the MF and $\mathbf{u}$ is the flow
speed. Rewriting the cross product in Eq.~\ref{vecprod} in terms of
dot products gives

\begin{equation}
\frac{\partial \mathbf{B}}{\partial t} = 
\mathbf{u} (\bigtriangledown \cdot \mathbf{B})-
\mathbf{B}(\bigtriangledown \cdot \mathbf{u}) +
(\mathbf{B} \cdot \bigtriangledown) \mathbf{u}-
(\mathbf{u} \cdot \bigtriangledown) \mathbf{B}.
\label{dotprod}
\end{equation}
Since we have spherically symmetric flow, $u_r = u_r$, $u_{\theta} =
0$, $u_{\phi} = 0$ and $\partial u_r/ \partial \theta= 0$, $\partial
u_r/ \partial \phi = 0$. Setting $\theta=\pi/2$, i.e., $B_\theta$ and $B_\phi$
scale in the same fashion and together form the tangential field, and using
\begin{equation}
\bigtriangledown \cdot \mathbf{B} = 0=
\frac{1}{r^2}\frac{\partial(r^2\,B_r)}{r} +\frac{1}{r}\frac{\partial B_\theta}{\partial\theta}
+\frac{1}{r}\frac{\partial B_\phi}{\partial\phi},
\end{equation}
we reduce Eq.~\ref{dotprod}
(see appendix) to a set of expressions that describe the evolution of
the radial and tangential components of $\mathbf{B}$,
\begin{equation}
\begin{array}{ll}
\frac{\partial B_r}{\partial t} = - 
\frac{\partial}{\partial r} \left(B_r u_r\right) -
2 B_r \frac{u_r}{r} +
B_r \frac{\partial u_r}{\partial r} \\
\frac{\partial B_{\theta}}{\partial t} = - 
\frac{\partial}{\partial r} \left(B_{\theta} u_r\right) -
B_{\theta} \frac{u_r}{r}\\
\frac{\partial B_{\phi}}{\partial t} = -
\frac{\partial}{\partial r} \left(B_{\phi} u_r\right) -
B_{\phi} \frac{u_r}{r},\\
\end{array}
\label{BrBt}
\end{equation}
Eq.~\ref{BrBt} represents the transport of the MF components with the
flow. To solve them, we need to apply initial values and boundary conditions.

In the core-collapse Type-IIb SN~1993J, the MF in the emission region is found to be
$\simeq64$~G at an age of a few days, when the SNR has a radius of roughly $R_{\rm SN} =
10^{15}$~cm \citep{FraBjo98, Maretal11}. Given the resolution of the radio data, it is not clear whether
the emission region in such a young SN is decoupled from the ejecta or not. Therefore we interpret
this MF as the average field of the SN and valid also for the ejecta. There are several arguments 
supporting this interpretation.
(i) During the collapse of a massive star, a number of processes such as magneto-rotational 
\citep{AkiWhe03} and stationary-accretion-shock \citep{Endetal10} instabilities, small-scale 
dynamo action \citep{ThoDun93}, Alfv\'en-wave amplification \citep{Guiletal11}, or simple winding 
of MF lines in differential rotation may result in a significant amplification of the MF 
\citep{ObeJan11}. The amplified field may be ejected outward together with part of the  
stellar material in the supernova explosion \citep{Endetal10, Endetal12, ObeJan11}. 
The expansion-diluted MF will be roughly as strong as observed in SN~1993J. 
(ii) Common sense suggests that the outflow of ejecta is not laminar. 
Therefore, various turbulence-induced MF-amplification processes \citep{Beretal09, 
Guoetal12, Sanetal12} may operate also at the RS propagating through the ejecta. 
Alternatively, it may be that the high observed MF is amplified by CR-streaming 
instabilities. In the latter case, however, inserting parameter values found by \citet{FraBjo98, Maretal11} 
into formulas for the saturation level \citep{Capetal09b, LuoMel09} yields values of the MF strength 
that are between one (non-resonant mode) and two (resonant mode) orders of magnitude lower 
than observed. Additionally, the time evolution of MF will not be as observed \citep{Dwaetal12}.

To ensure that initially $\bigtriangledown \cdot \mathbf{B} = 0$ in the
ejecta, we assume that $B_{\rm ej,r}$ is flat where $u_r \le u_{fl}$ and $B_{\rm ej,r}(r)\propto 1/r^2$ where $u_r > u_{fl}$. We assume the same for the tangential field,
and scale all components so that the total volume-averaged MF strength is $\simeq 50$~G
when $R_{\rm SN} = 10^{15}$~cm. For the initial condition at $t_0=1$~year, we use the MF scaled according to flux conservation:
\begin{equation}
B_i(r,t_0) = B_{\rm ej}(r) \frac{R_{\rm SN}^2}{R_{\rm SNR}(t_0)^2}.
\label{icon}
\end{equation}

In the current work we limit ourselves to scenarios with no
amplification of the MF upstream of the FS, and only the 
CSM field is assumed to be
transported through the shock to the downstream region. The 
MF in the circumstellar medium (CSM) 
of Type-Ic and Type-IIP SNRs is shaped by the winds of the progenitor
stars. We assume that turbulence equalizes all 
components of the MF in the wind zones until they are transported through the FS.
The profile of the MF in the environment
of a core-collapse SNR in our description is then given by
\begin{equation}
B(r) = \left\{ \begin{array}{ll}
      B_{\rm b}(r)		& \textrm{$R_{\rm st} \leq r \leq R_{\rm b}$}\\
      B_{\rm b}(R_{\rm b})\sqrt{11}	& \textrm{$R_{\rm b} \le r \leq R_{\rm sh}$},
\end{array} \right.
\label{CSMF}
\end{equation}
where $B_{\rm b}(r) = B_{\rm st}R_{\rm st}/r$, $B_{\rm st}$ is the MF
at the surface of the progenitor star ($\approx 100$~G for WR stars,
$\approx 1$~G for RSGs), $R_{\rm st}$ is the radius of the progenitor
star ($\approx 8R_{\odot}$ for WR star, $\approx 600R_{\odot}$ for
RSG), $R_{\rm b}$ is the radius of the wind-blown bubble in our
simulations ($\approx 7$~pc for WR star, $\approx 2$~pc for RSG),
$R_{\rm sh}$ is the radius of the swept-up shell ($\approx 30$~pc in
both cases), and $r$ is the distance from the star. Note that the $1/r$ scaling of the field strength results from the advection of the stellar MF in a wind, colloquially known as the Parker spiral, 
where it applies to the tangential components only. The radial field in the wind zone
is supposed to arise from turbulence, which may in fact also amplify 
it \citep{DruDow12}, and should be understood as RMS amplitude. 
Magnetic-field amplification by cosmic-ray streaming \citep{2000MNRAS.314...65L}, 
upstream dynamo action \citep{2009ApJ...707.1541B}, the Richtmeyer-Meshkov 
instability \citep{Sanetal12}, or vorticity-generation at shocks \citep
{2007ApJ...663L..41G} are not considered here.

The solution of Eq.~\ref{BrBt} is obtained numerically, using
the initial conditions
described above and the parameters of the
SNR plasma flow from the simulations described in
Section~\ref{sec:hd}. The time-dependent solution for different SNR
types is plotted in Fig.~\ref{mf}. Our solutions for the MF evolution
are consistent with earlier 1-D studies \citep{RosFra76}, but 2-D MHD
solutions \citep{Schetal09} show significant distortions at the
contact discontinuity due to the growth of the Rayleigh-Taylor (or Kruskal-Schwarzschild)
instability.

\section{Particle acceleration method}

Our method is based on time-dependent kinetic calculations in a
test-particle approximation \citep{Teletal12a, Teletal12b}. They are implemented by
numerical solution of the diffusion-advection equation for the differential
particle number density on a grid co-moving with the shock wave and in
spherically symmetric geometry. To resolve the diffusion
length of the lowest energy particles, we performed a coordinate
transformation that increases the spatial resolution near the shock. The
spatial coordinate, $x$, is related to the new coordinate, $x_*$,
for which a uniform grid is used, by the equation
\begin{equation}
(x-1)=\left(\frac{r}{R_{\rm SH}} -1\right)=(x_*-1)^3\ ,
\label{transform}
\end{equation}
where $R_{\rm SH}$ is the shock radius. Thus, with modest resolution in
the $x_*$, one may achieve a very fine resolution in $x$ where it matters,
namely in the shock region where the newly injected
low-momentum particles enter the acceleration process. The other
benefit is the significant extension of the grid toward $x\gg 1$ at
very low computation cost, since the range in $x$ is proportional to
the third power of $x_*$. The boundary condition for large $x$ then
does not really affect our solution, since particles do not
leak out of the grid but rather are distributed over the huge upstream
volume according to the diffusion properties of the media. We included
an exponential transition from Bohm to Galactic diffusion
at around $2R_{\rm FS}$. A high Galactic diffusion coefficient also compensates
for the deteriorating resolution of the spatial grid
beyond $2R_{\rm FS}$. Because our method is based on the test-particle
approximation, we did not permit the CR pressure at the shock to exceed 10\% of
the ram pressure. We used a thermal-leakage injection model
\citep{Blaetal05}, and adjusted the injection so that the CR-pressure
limit was not violated. The injection coefficient, a free parameter, was
taken to be approximately $5\cdot10^{-6}$ for Type-Ic and for
$5\cdot10^{-8}$ Type-IIP SNR.

\section{Results and discussion}

Based on the hydrodynamic simulations described in
Section~\ref{sec:hd} and the MF profiles obtained in
Section~\ref{sec:mf}, we compute time-dependent particle distributions
accelerated by the forward and reverse shocks. We calculate the
resulting emission due to synchrotron, inverse Compton, and pion-decay
processes, build corresponding surface-brightness maps and discuss the
observational implications. We present snapshots at the age of 400,
1000, and 1975~years to trace spectral and morphological
evolution of emission coming from Type-Ic and Type-IIP SNRs.

\begin{figure*}[!t]
\vspace{10mm}
\centering
\includegraphics[width=0.99\textwidth]{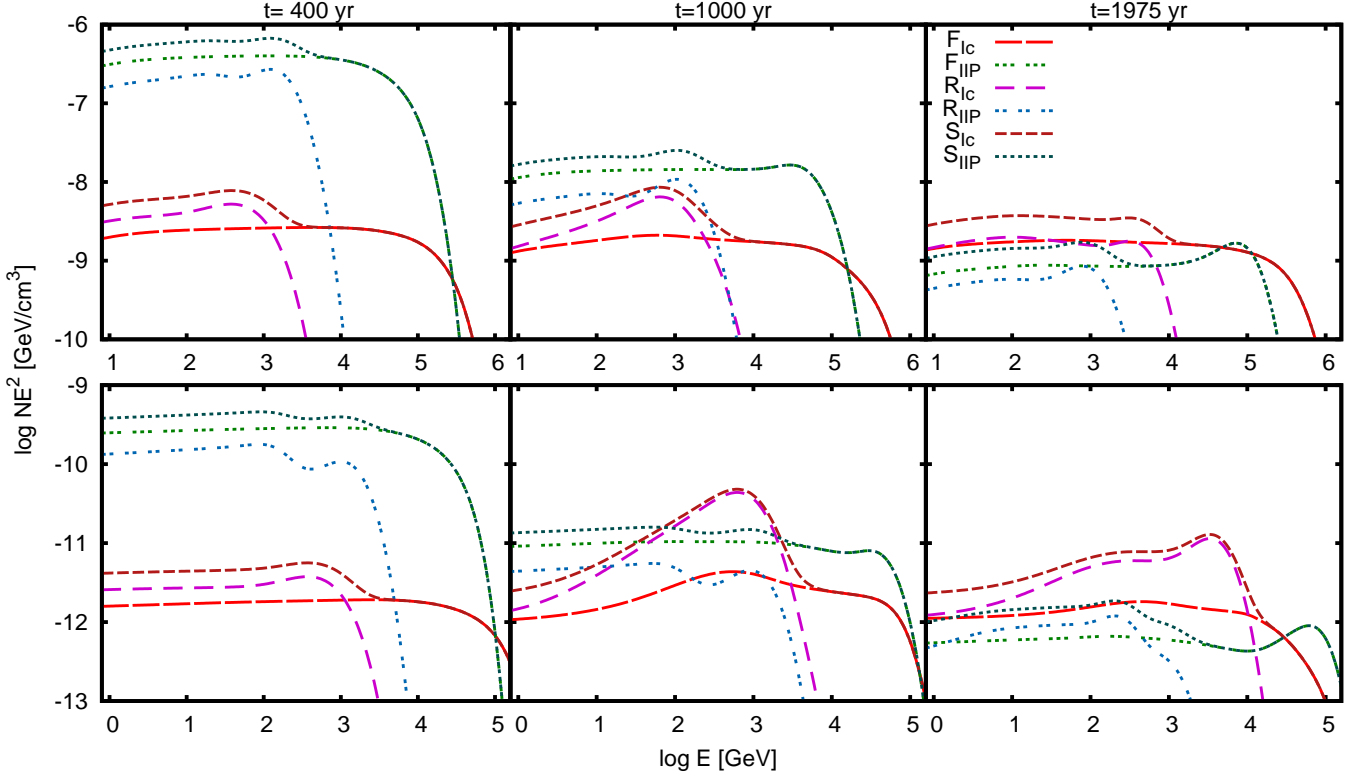}
\caption{Time evolution of the proton (top) and electron (bottom) spectra
for Type-Ic and Type-IIP SNR. Total volume-integrated spectra (S) are 
the sum of the contributions from the forward (F) and reverse (R)
shocks.}
\label{spe}
\end{figure*}

\subsection{Particle spectra}
\label{sec:spe}
Particle spectra for Type-Ic and Type-IIP SNRs are given in Fig.~\ref{spe}.
We show spectra produced by acceleration at the forward and reverse shocks
plotted along with a volume-integrated total spectrum from the whole SNR.

At an age of 400~years, the Type-Ic SNR is still propagating through
the freely expanding stellar wind. The flow velocity and density
profiles are given by the classical self-similar solutions. The
spectrum of protons accelerated by the FS is well described by a
power-law with exponential cutoff, in agreement with classical DSA
solutions. The magnetic field at the FS is still strong at this age, and so
particles reach high energies. The spectrum of RS-accelerated protons is
also consistent with standard DSA. Note that the volume integration 
was performed over both the upstream and the downstream region of the RS,
whereas the upstream region of the FS was not covered 
(for details see Section 5.1.1 of our previous paper \citep{Teletal12a}). The magnetic
field at the RS is considerably smaller than at the FS, and so particle
energies reach a TeV only. However, the number of
particles injected into the acceleration process, and hence the density, is
larger. Therefore, below a few TeV the CR number density at the RS
is higher than that at the FS, and the total volume-integrated proton spectrum displays a
downward step around a few TeV. In phenomenological studies of SNRs
such spectra may be interpreted as the sum of different populations of particles coming from different zones of
the remnant \citep{AraCui10, AtoDer12, Lemetal12}. Since the MF is not 
strong enough at either of the shocks to cause significant
synchrotron losses, the electron spectra look similar to those of protons.

The Type-IIP SNR at the same age also evolves in the
freely expanding progenitor wind, though this wind
has a much higher density. Therefore both electron and proton spectra
have a much higher intensity than in the Type-Ic SNR. The density
at the RS of the Type-IIP SNR is lower than at the FS, and consequently the CR
intensity is lower. Nonetheless, the contribution of the RS to the
overall particle spectrum is noticeable. The magnetic field at the RS
of the Type-IIP SNR is slightly larger than in the Type-Ic case
because the radius is smaller, and particles are accelerated to
somewhat higher energies, cutting off at around a few TeV. For the same
reason, electron losses in the downstream region of the RS are more
efficient and steepen the electron spectra beyond a characteristic
energy, $E_{\rm rad}$, downstream of the RS, whereas in the upstream region
of the RS the MF, and hence the electron energy losses, are weak. Since 
the volume-integrated spectrum of particles produced by the RS includes both
the downstream and the upstream contributions, we observe a bump located at
$E_{\rm max}$ of the RS upstream electron distribution, which is located
beyond $E_{\rm rad}$ (see Section 5.1.1 of \citet{Teletal12a} for
details).

At 1000~years, Type-Ic SNR particle spectra produced by the FS differ
significantly from those expected via standard DSA. The 
significant spectral modification can be traced back to the 
interaction of the FS with the wind termination shock, at around
700~years. The effective compression ratio at the FS as seen by the CRs increased
significantly. The acceleration was boosted for a short time, and 
since the interaction time was brief and the spectral
index hard, a bump at a few hundred GeV formed, which slowly moved
toward higher energies as the particles in it were
further accelerated by the FS. 

\begin{figure*}[!t]
\vspace{5mm}
\centering
\includegraphics[width=0.98\textwidth]{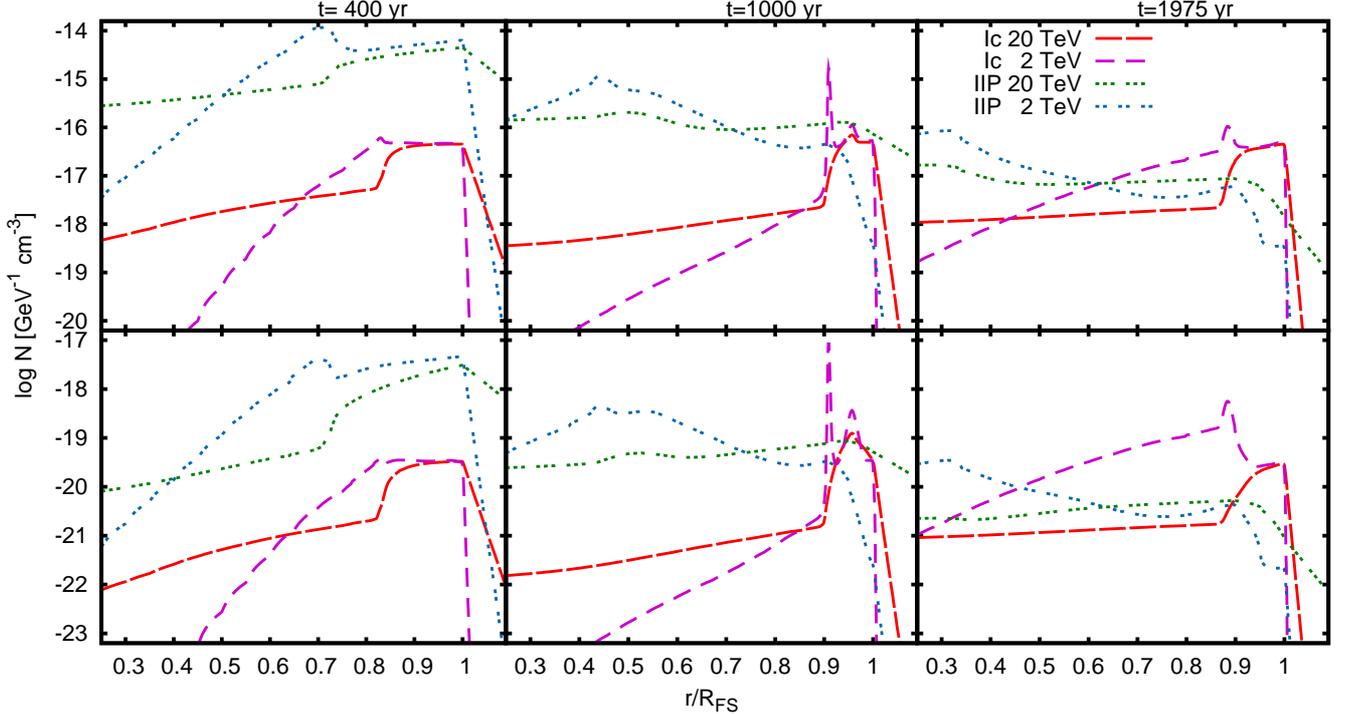}
\caption{Time evolution of the radial distribution of protons (top) and
electrons (bottom) at an energy of 2~TeV and 20~TeV for different SNR
types and models. The 2-TeV distributions were downscaled by a
factor 100 for better visualization.}
\label{prof}
\end{figure*}

By the time of 1000~years, a break at around 1~TeV is still
clearly visible. When the FS collided with the wind termination shock,
a reflected shock that propagates inward formed. By coincidence, the
time when the reflected shock reaches the RS of the remnant is very
close to 1000~years. So the RS spectra shown are rather specific for
this particular time, but we show them to illustrate the impact of
the hydrodynamics on particle spectra. In analogy with the FS, the
compression ratio at the RS increased for a very brief time of
interaction, the spectral index became very hard, and a bump at a few
tens of GeV formed and shifted to a few hundred GeV by the time of
1000~years.

Compared to the Type-Ic, Type-IIP SNRs do not show such a strong
variation in particle spectra at the same age. The forward shock of
the Type-IIP SNR encountered the MS wind zone at around
600~years. Since the shock entered a very dilute environment, no
significant additional compression occurred at the FS, but rather a
very brief de-compression that did not affect CR spectra. After
  the encounter, another reverse shock formed and propagated through
  the FS-compressed CSM, and so we see a three-shock structure: the
FS propagating through the MS wind zone, the internal RS going through
the ejecta, and another reverse shock between them propagating through
the dense CSM compressed earlier by the FS. Although the divergence of
the plasma flow velocity is negative between the forward and
intermediate shock, thus permitting re-acceleration of particles
  in this region and at the intermediate shock itself, we cannot see
the effect in volume-integrated spectra. The FS speed increased
when it entered the diluted-wind zone. The maximum energy of CRs,
  $E_{max}$, increased as expected, but the rate of particle
  injection dropped significantly. The separation between the forward
and reverse shocks started to increase on account of the boosted
velocity of the FS, and so the volume enclosed by the RS
  constitutes a diminishing fraction of the total SNR
volume. Nevertheless, the contribution of the RS-accelerated particles
remains visible because the RS propagates through very dense ejecta.

Nearly 1000~years later, at the age of 1975~years, the velocity
profiles of both remnant types have relaxed. The multiple weak reflected
shocks in Type-Ic SNR have dissipated, but the CR spectra still
retain their effects, though in general they
appear smoother than before. The proton distributions from both the
FS and the RS are softer, and the contribution from the RS to
the total proton spectra remains visible. The electrons accelerated by
the FS have rather soft spectra as well, while those
accelerated by the RS dominate the total electron spectra of Type-Ic SNR.

At an age of 1975~years, we see an excess of 
high-energy particles accelerated by the FS in the Type-IIP SNR.
The electron spectra differ only slightly from the 
proton spectra due to weak synchrotron losses. Once the FS entered the 
dilute zone, the particle injection rate dropped four orders of magnitude.
It takes time for the particle spectra to adjust to the new equilibrium, 
particularly so at very high energies. 
For the high-energy particles the characteristic diffusion-advection timescale,
$\tau_{\rm da} = \kappa/v_s^2$, is long, and therefore they continue to diffuse around 
the shock and gain energy while the lower-energy particles were advected from the 
shock. We can consider the bump as a signature of the delayed reaction of the 
high-energy particles to a sudden change in the injection rate.

\subsection{Radial distributions}

The radial distributions of 2-TeV and 20-TeV particles are presented
at Fig.~\ref{prof}. There is a clear difference between particle
profiles depending on both energy and SNR type.

The 20-TeV particle profiles comprise only particles accelerated at
the FS, because the RS does not accelerate CRs up to this energy on
account of the small MF. On the other hand, the RS region is prominent
in the 2-TeV particle profiles because at this energy CRs do not
efficiently propagate away from their acceleration sites. Since
cosmic-ray electrons lose energy much faster than protons, their
intensity falls very quickly with increasing distance from the shocks.

At the age of 400~years, Type-Ic SNRs show an increased number density
of 20-TeV particles between the two shocks. There, the MF is higher
than in other parts of the SNR, and so particles are retained.
In the ejecta region, where the MF is low, particles are distributed
rather uniformly with an intensity approximately an order lower than
in the shocked region. The intensity of 2-TeV particles shows an
enhancement in the region of the RS on account of the RS acceleration,
but only marginally so because the cutoff energy of particles
accelerated at the RS is below 2~TeV. A steep decrease of intensity
toward the SNR center arises because the diffusion time of 2-TeV
particles is long and the age of the SNR is insufficient to
distribute 2-TeV particles uniformly.

\begin{figure*}[!t]
\vspace{10mm}
\centering
\includegraphics[width=0.99\textwidth]{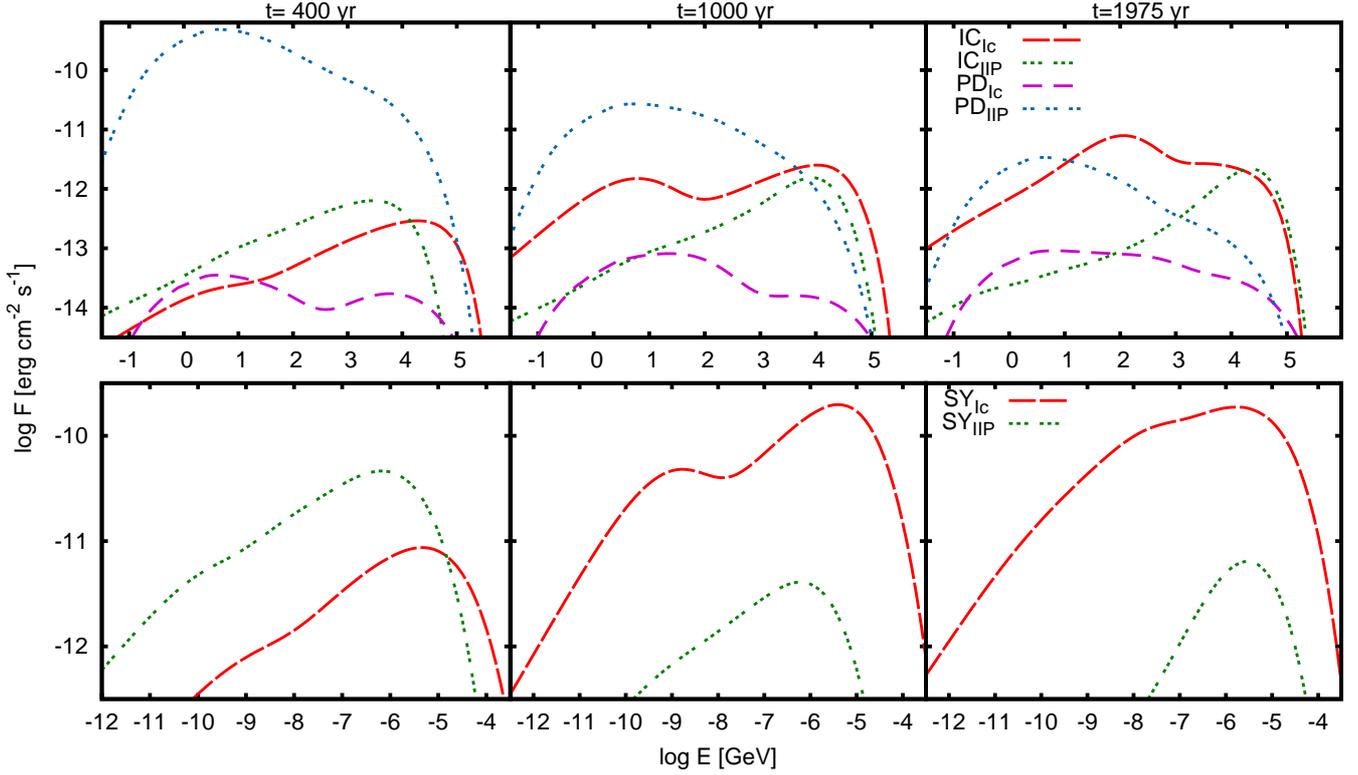}
\caption{Time evolution of the emission spectra from Type-Ic and Type-IIP SNRs due
  to pion-decay (PD), inverse Compton (IC), and synchrotron (SY)
  radiation.}
\label{rad}
\end{figure*}

The profiles of 20-TeV particles in Type-IIP SNRs of the same age
behave similarly.  The intensity of 2-TeV CRs shows a significant peak
at the RS, though, because the MF at the reverse shock of Type-IIP SNR
is stronger than in Type-Ic SNR. The maximum energy of RS-accelerated
particles is therefore considerably higher than in the case of Type-Ic
SNRs, in fact it is well beyond 2~TeV, thus leading to a prominent
contribution of CRs from the RS.

As we mentioned above, the age of 1000~years happens to be a
special phase in the evolution of Type-Ic SNR for our choice of
parameters, and it is interesting to see what happens to particle
distribution just after the shocks collide. The profiles of 20-TeV
particle are largely unaffected by the recent interaction. The MF at
the RS is insufficient to re-accelerate particles up to 20~TeV and
raise the intensity at this energy. A small peak in between two shocks
can be explained by the increased MF at the contact discontinuity
behind one of the weak reflected shocks; a similar structure is
visible in the profiles of 2-TeV particles. This MF creates a barrier
for particle diffusion to the SNR center. The 2-TeV CR profiles are
strongly affected by the recent shock collision. An excess of 2-TeV
particles is clearly seen at the RS and, in fact, also in the
volume-integrated particle spectra. Since the particles have no time
to diffuse around the acceleration region, the peak is very sharp.

After 1000 years, the radial distribution of CRs in Type-IIP SNR has
changed dramatically compared to a few hundred years earlier when the
FS entered the MS wind zone. At the age of 1000~years the bulk of the
CRs is relatively old, having been accelerated when the FS was
propagating through the dense RSG wind. The high intensity of CRs,
which were advected inside the SNR, joins smoothly with the low
intensity of particles freshly accelerated by the FS. The profiles of
20-TeV particles are significantly smoothed on account of fast
diffusion, whereas the intensity of 2-TeV particles shows an 
increase toward the RS. The RS propagating through dense ejecta at
$0.4 R_{\rm FS}$ strongly contributes to the profile of 2-TeV
particles.

After 1975~years, the particle profiles in Type-Ic SNR have lost
memory of the reflected shocks present earlier. The spatial
distribution is rather smooth with enhanced intensity in the shocked
region, at 2~TeV particularly so at the RS suggesting that
acceleration is still ongoing. Low-energy particles are far more
abundant in the SNR interior than they were at early times. At 20~TeV
the intensity is nearly constant in the SNR interior.

The radial profiles of particles in Type-IIP SNR have not changed much
from 1000 to 1975~years. The profiles became smoother, but the peak
at the RS at around $0.3 R_{\rm FS}$ is still observed for 2-TeV
particles. Interestingly, the intermediate shock (see previous
subsection) is now visible as an increase in low-energy particle
intensity at around $0.9 R_{\rm FS}$. Although no impact of the
intermediate shock is seen in volume-integrated spectra, it clearly
affects radial profiles, which suggests that some re-acceleration occurs 
between the forward and intermediate shocks in addition to particle
  trapping at the contact discontinuity. This effect is not as
relevant for 20-TeV particles on account of fast diffusion, and so
their distribution is rather smooth.

\subsection{Nonthermal emission}

\begin{figure*}[!t]
\vspace{5mm}
\centering
\includegraphics[width=0.32\textwidth]{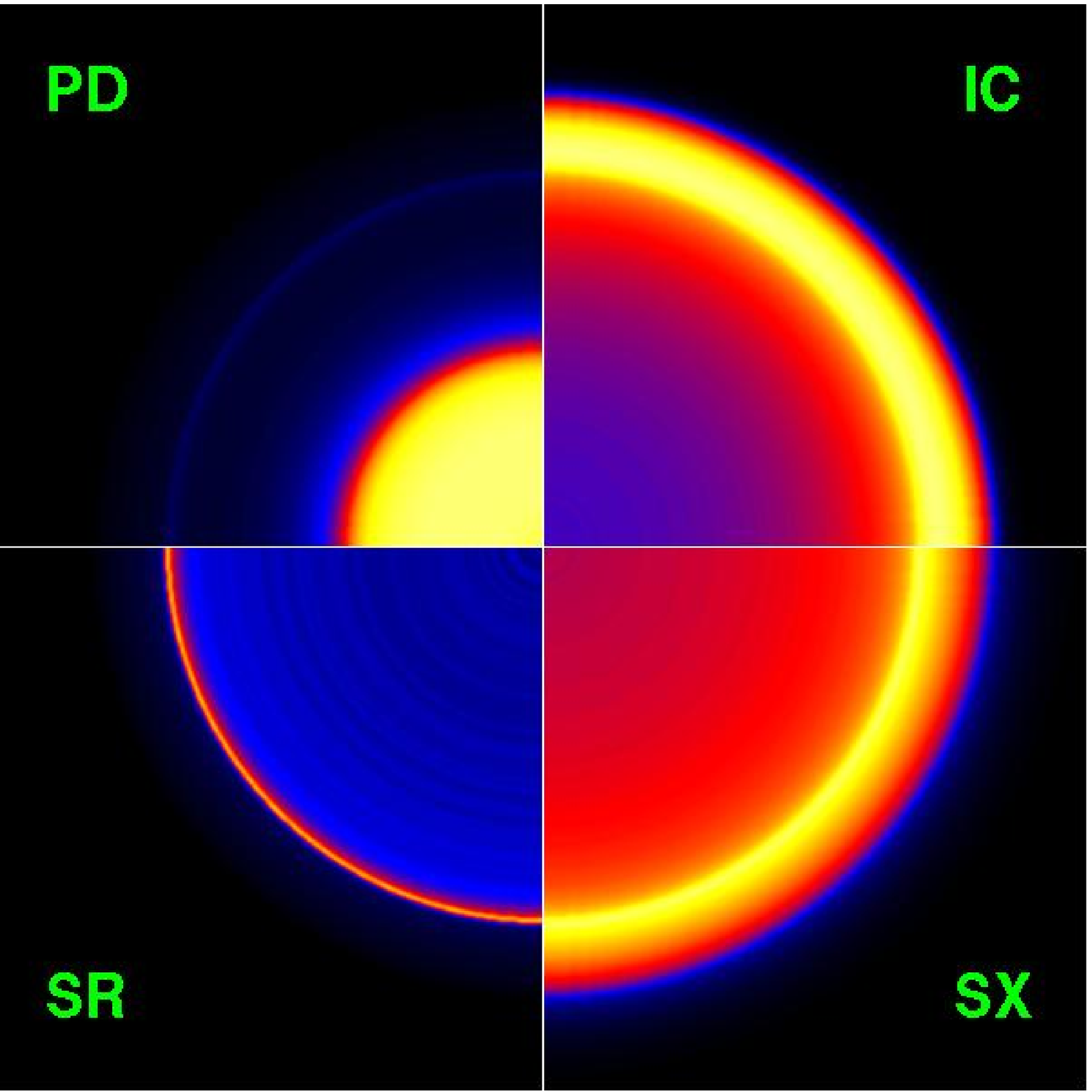}
\includegraphics[width=0.32\textwidth]{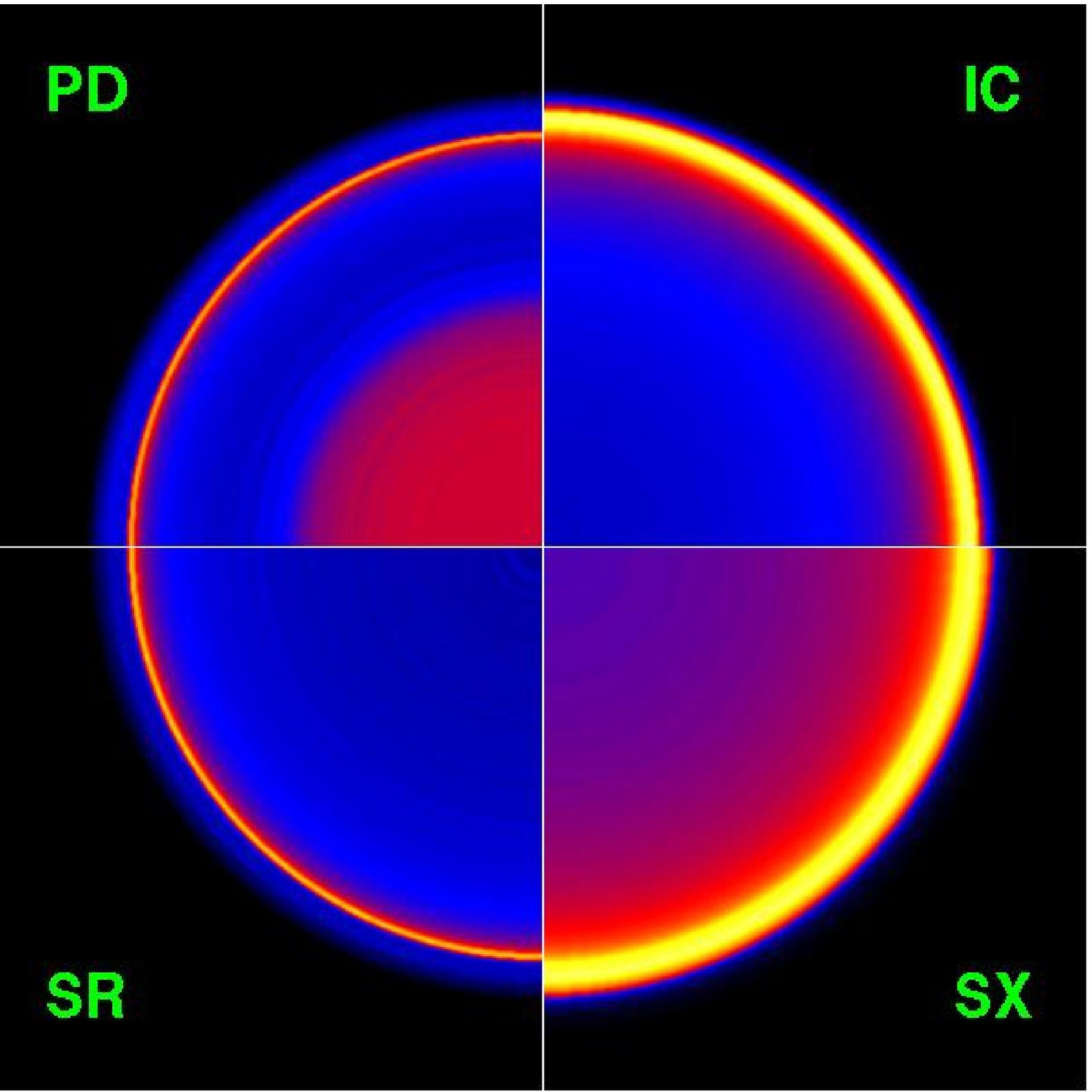}
\includegraphics[width=0.32\textwidth]{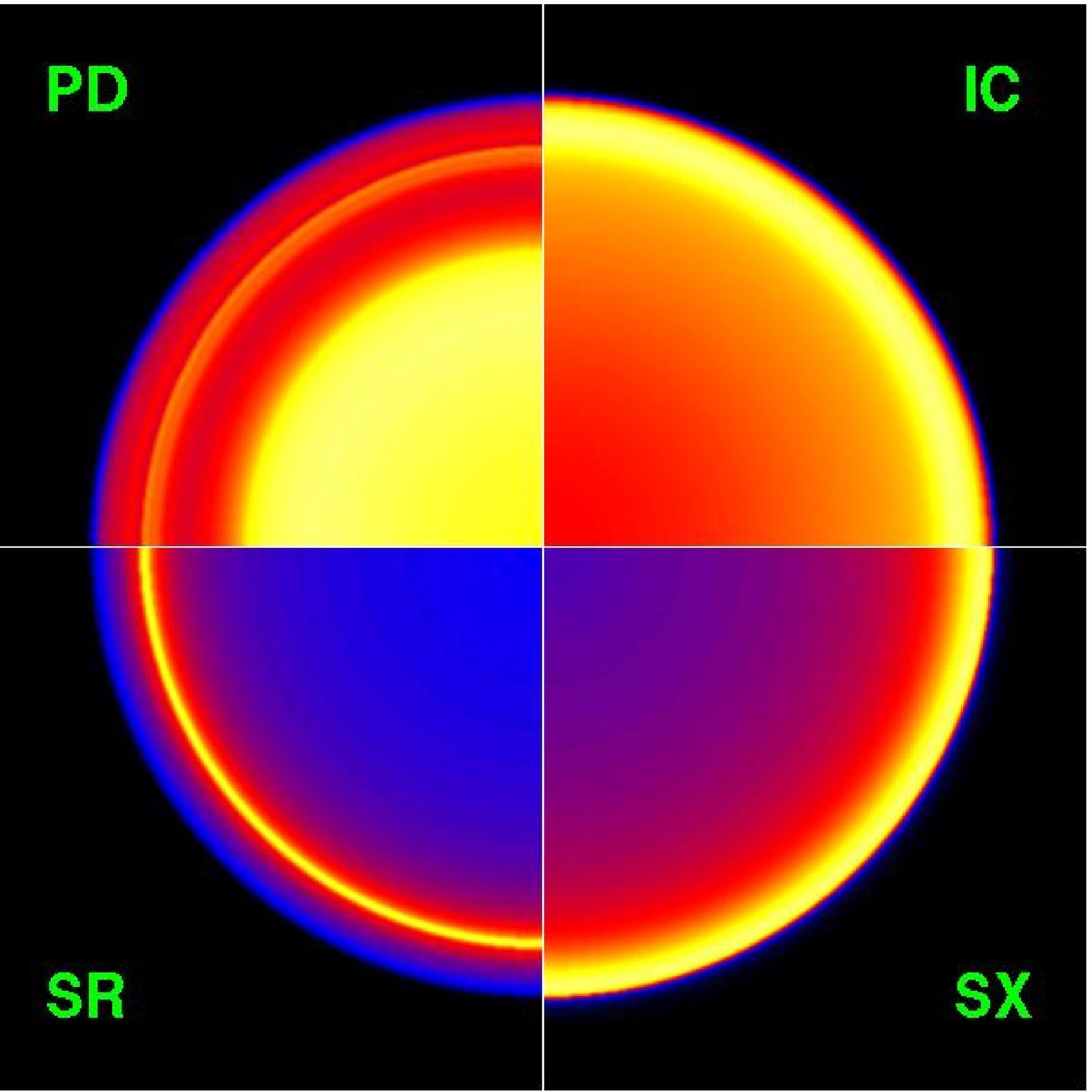}
\includegraphics[width=0.32\textwidth]{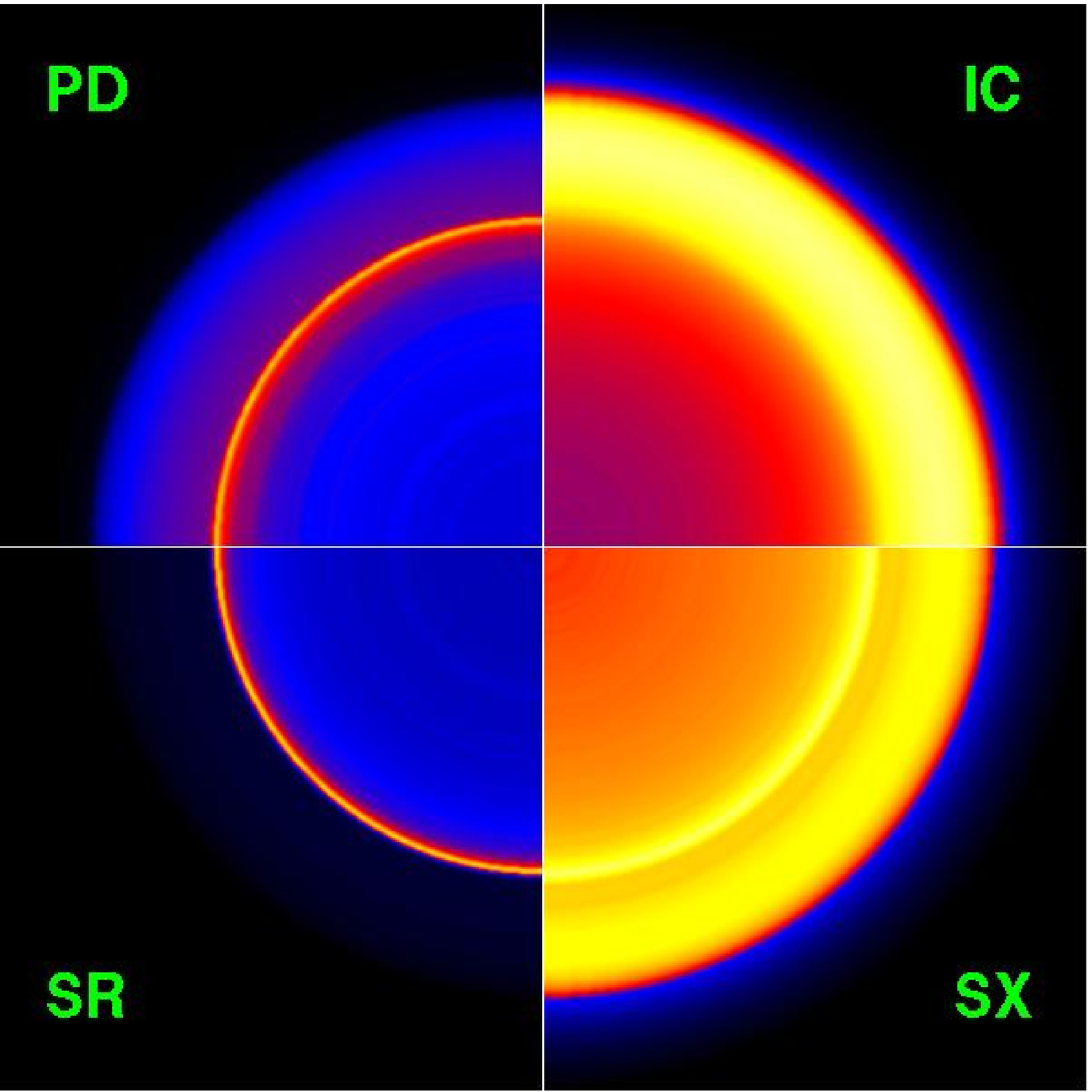}
\includegraphics[width=0.32\textwidth]{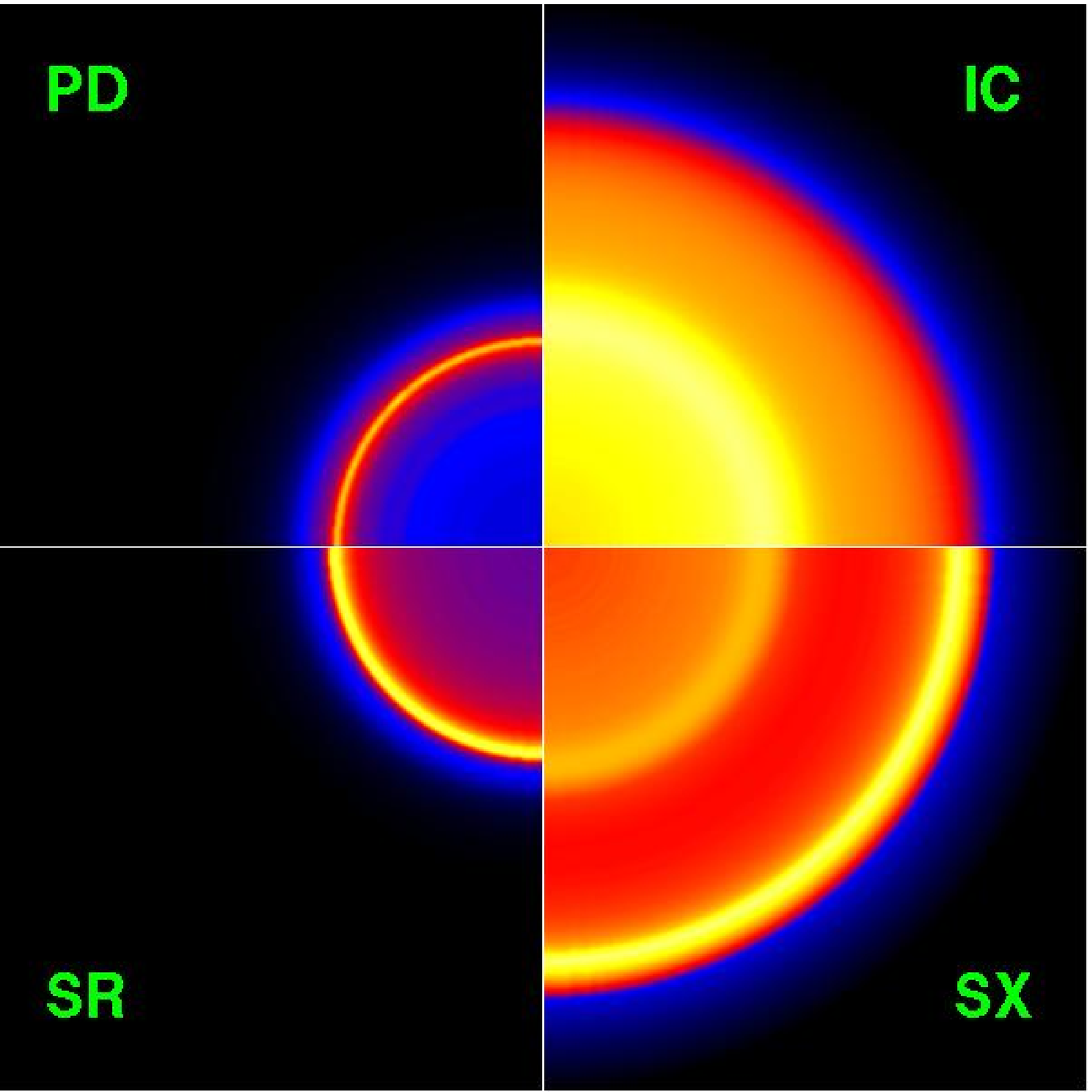}
\includegraphics[width=0.32\textwidth]{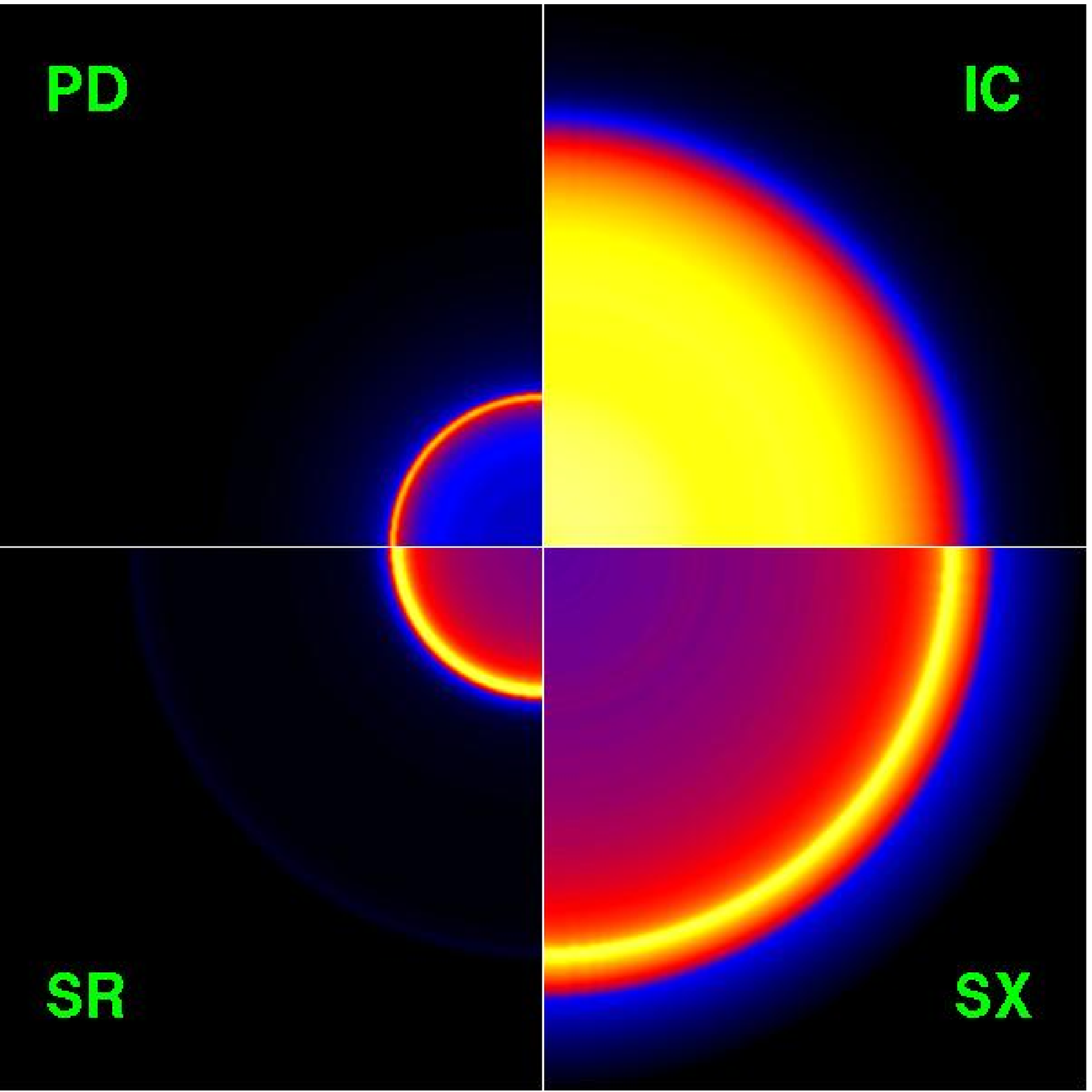}
\caption{Time evolution of intensity maps of Type-Ic (top) and
  Type-IIP (bottom) SNRs at 1~TeV due to pion-decay (PD) and inverse
  Compton (IC), at 3~keV (SX) and at 1.4GHz (SR) due to synchrotron
  radiation. The left column is for the age of 400~years, the middle column at 1000~years, and
the right column at 1975~years. The scale is linear from zero to maximum in each image. All images are normalized by the FS radius, $R_{FS}$, at the respective time.}
\label{maps}
\end{figure*}

We considered three radiation processes of non-thermal particles:
synchrotron and inverse Compton (IC) emission of electrons, and
neutral pion decays originating in collisions of CR protons with
protons at rest. Only primary electrons are considered for the
leptonic processes. The calculated photon distributions for Type-Ic SNR and
Type-IIP SNR are plotted at Fig.~\ref{rad}. We also calculate, and discuss in context with the
radiation spectra, intensity maps for the emission at characteristic
radio, X-ray and gamma-ray wavelengths. The maps are shown in
Fig.~\ref{maps}. We note that in all calculations of volume-integrated
spectra or intensity maps the radial distributions of CRs, the target
material, and MF are taken into account. The result is
significantly different from simple toy-model calculations assuming
uniform particle, density, and MF profiles inside the SNR. We first
give brief details on the method of calculation and continue with a
discussion of our results.

The relevant MF component for synchrotron emission is that in the
plane-of-sky, $B_{\perp}$, which is perpendicular to the line of sight
(LOS). If all components of the MF are equal, then it is
straightforward to show that the effective value of $B_{\perp} \simeq
0.8 B$. When properly accounting for the transport of the MF to the
downstream region of the FS, it is clear that the radial and the
tangential components are not equal. We therefore derive the
plane-of-sky component of the MF as
\begin{equation}
B_{\perp} = \sqrt{B_t^2 sin^2 \xi cos^2 \alpha + B_t^2 cos^2 \xi +
  B_r^2 sin^2 \alpha + B_t B_r \sin\xi \sin(2\alpha)},
\end{equation}
where $B_r$ is the radial component of the MF, $B_t =
\sqrt{B_{\theta}^2 + B_{\phi}^2}$ is the tangential component of the
MF, $\alpha$ is the angle between the radial direction and the LOS,
and $\xi$ is the angle between the direction of the tangential field
and the $z$-axis. Since $\xi$ is unknown and in general may assume any
value in the range [0,2$\pi$], there are two options to choose it. One
option is to keep it constant at $\xi=\pi/4$, the other one is to
randomly vary it. The latter case would assume strong turbulence and
little radio polarization \citep{2009ApJ...696.1864S}.  Here we chose
$\xi=\pi/4$ and applied the standard formula for synchrotron emission as
for instance used by \citet{Stuetal97}.

The IC emission was calculated using the full Klein-Nishina cross
section \citep{BluGou70} for relativistic electrons following
\citet{Stuetal97}. As the target photon field for the IC scattering we
considered the microwave background only. We calculated hadronic
gamma-ray emission according to the procedure described by
\citet{Huaetal07}.

Thermal emission was not considered here. However, we estimated this
given the SNR parameters under consideration. Thermal emission should be very bright for young
Type-IIP SNR on account of the high gas density in the RSG wind.
However, at very early times Type-IIP SNe tend to show the lowest X-ray intensity
of all SN types \citep[see Fig.~3 of][]{DwaGru12}. The reason may
be that the high density also provides strong absorption. 

Most of the 1-TeV IC and pion-decay emission is created by particles
of roughly 20-TeV energy, whereas the energy of electrons producing
1.4-GHz and 3-keV synchrotron maps depends on the MF, which varies
with age, location within the SNR, and type of SNR. The general trend
to be noted from Fig.~\ref{rad} is that the high-energy emission
from Type-Ic SNRs is dominated by IC emission, with a non-negligible
contribution from pion-decay only at the early age of 400~years.
Fig.~\ref{rad} shows that the high-energy emission of Type-IIP SNR
is mostly hadronic, with a non-negligible contribution from IC only at
the late age of 1975~years.  This is not surprising, because the very
thick RSG wind provides a good target for hadronic CR interactions in
Type-IIP SNR, whereas Type-Ic SNRs expand in the low-density wind of
the WR star. While the high-energy flux from Type-Ic SNRs is
increasing with time, that from Type-IIP SNRs is decreasing. Type-IIP
SNRs are always brighter in gamma-rays than Type-Ic remnants with the
exception of late times when the fluxes are of the same order. The
synchrotron emission of both SNR types shows the opposite trend. To be
noted from Fig.~\ref{rad}--\ref{maps} is that the spectra and
intensity maps have a few distinct particle populations and emitting
zones. Thus complex spectra and morphology arise naturally, but evolve in time.

At the age of 400~years, the hadronic emission from Type-Ic SNRs comes
from two regions: the region of dense ejecta provides a high-energy
bump in the spectrum, and the region around the contact discontinuity
accounts for a low-energy bump. The energetic CRs accelerated at the
FS penetrate deep into the SNR and illuminate the ejecta, while
low-energy CRs are trapped around the RS and illuminate gas at the
nearby contact discontinuity. The IC emission at this time comes
mostly from high-energy electrons of the FS.  The synchrotron emission
extends to hard X-rays. It comes mostly from the region between the FS
and the contact discontinuity, somewhat increasing in intensity
toward the CD on account of increasing MF
\citep{RosFra76,2004ApJ...609..785L}. The radio emission is clearly
dominated by the emission from the CD, where both MF and low-energy
electrons are in excess.  We point out that the decelerating CD is
subject to the Rayleigh-Taylor instability, which we cannot treat in
our 1-D calculations. Therefore, real brightness maps would show a
considerably more patchy structure near the contact discontinuity. The
idealized picture presented here should nonetheless trace the
general trend of the enhanced emission near the CD of core-collapse
SNRs.

The Type-IIP SNR at the same age shows quite different spectra and
morphology of emission. Hadronic emission comes from the entire
remnant with a significant contribution from the CD. The PD spectrum
is therefore very soft because the emission region is dominated by the
low-energy CR population accelerated at the RS. It is worth noting 
that the turnover in the
PD spectrum is around 10~GeV, a feature observed in other core-collapse remnants
as well \citep[e.g. Cas~A][]{Accetal10_CasA,
Abdetal10_CasA}. The IC emission is produced in the region between
the CD and the FS, but its intensity is low. The whole SNR is very bright in
synchrotron X-rays because the MF is high throughout. This
makes it possible to clearly distinguish the RS, the CD, and the FS.
The radio emission, as in Type-Ic case, is dominated by the
CD region.

At the age of 1000~years, the gamma-ray emission from Type-Ic SNR is
dominated by IC radiation. The spectrum of the emission is double-peaked: at low
energies it is dominated by the particles recently accelerated in
the shock collision, whereas the high-energy IC emission is produced
by particles accelerated at the FS.
The PD spectrum is of low intensity with a significant bump at low energies.
The IC and PD morphology maps at 1~TeV are dominated by high-energy CRs, and
therefore no hint of emission from RS-accelerated particles is seen. The 
PD emission of FS-accelerated protons illuminating the dense ejecta 
is still visible. The CD is now much brighter than the ejecta region, because the
target density at the CD is on the same order as in the ejecta, but the CR
density is much higher. The region between the CD and
the FS, where the CR density is high, is bright in IC emission. 
Since the MF in the
shocked region became approximately uniform, X-ray synchrotron
emission predominantly comes from the same region, thus
repeating the IC morphology. Likewise, the synchrotron
spectrum largely resembles that of IC emission, except for minor differences
on account of MF radial profile. The radio emission
predominantly arises at the CD, where the MF is higher than at the RS itself.

The very high gas density and fairly uniform distribution of CRs
in 1000-year old Type-IIP SNR renders the PD process dominant in the
emission spectrum. However, the emissivity is mostly localized in the
ejecta region. The CD is very bright, and the abundance of low-energy
protons there makes the PD spectrum soft. The IC emission
  originates mostly in the weakly magnetized ejecta and the region
  between the two contact discontinuities (one behind the reverse
  shock and the other behind the intermediate shock). Both CDs are very
  prominent in synchrotron X-rays. The radio synchrotron map is again
  dominated by the CD behind the reverse shock on account of the
  distribution of low-energy electrons.

At 1975 years of age, the high-energy radiation from Type-Ic SNRs
  is strongly dominated by IC scattering, with a significant
  contribution coming from low-energy electrons at the RS and ejecta
  regions. Most of the high-energy IC emission comes from the shocked region,
  where the electron number density is high. The PD radiation is very
  dim and arises mostly in the ejecta, with some contribution from the
  shocked region and the CD.  Synchrotron X-rays are emitted in the
  region between the FS and the CD, where both the number density of
  high-energy electrons and the MF are high. The radio emission is
  very bright around the CD and the RS, where the density peak of
  low-energy electrons is located and the MF is high. Since the ejecta
  have a very low MF, the low-energy peak in the synchrotron spectra has
  nearly disappeared. This demonstrates how the radial profile of the
  MF can affect the total emission spectrum, and that a simple scaling
  of the synchrotron emission to IC does not work.

The brightness of a 1975-year old Type-IIP SNR is still dominated by
the pion-decay emission coming from the compact ejecta region and the
bright CD, but only up to energies of around 1~TeV.  At higher
energies IC emission takes over, which arises throughout the remnant
out to the intermediate shock, predominantly tracing regions of low
MF.  In the region of strong MF, between the intermediate and forward
shock, we observe hard X-ray synchrotron emission. Radio emission is
significant in the ejecta region only. One can see that not only the
spectra of IC and synchrotron emission do not resemble each other, but
also their morphology is strikingly different. Thus care must be
  exercised in interpreting the observations if one attempts a
  reconstruction of the high-energy morphology from radio and X-ray
  synchrotron maps of core-collapse SNRs \citep{Petetal09}. A clear
  shell-type SNR in X-rays may appear as a compact structure in
  high-energy gamma-rays, which may be mistaken for a pulsar-wind
  nebula, thus confusing the identification of high-energy gamma-ray
  sources. Note that the very dense CD and ejecta have not expanded
  significantly, and the radius of the CD is only around 3~pc after
  1975~years.  Resolving this region is a challenge for the current
  TeV-band telescopes, because the angular diameter would be only
  $0.17^\circ$ at a distance of $2$~kpc.

\section{Conclusions}

Using realistic models for the circumstellar medium created by
  stellar mass-loss around massive stars, we have studied the
  evolution of the resulting SNR shock waves within this medium, and
  computed the very high-energy signatures from the same. Our
  calculations of the medium into which SNRs evolve took into account
  the changes in mass-loss rate as the progenitor star evolved through
  various stages. In this way, our calculations far more precisely identified
  the various structures in the surrounding medium than those of
  previous authors. For instance, \citet{Cap11} assumed a single model
  to describe the circumstellar medium around all massive stars, with
  an RSG wind region extending into a constant density WR region. Not
  only is such a single description unsuitable over the large
  parameter space of stellar masses, but the structure these authors have
  described is not consistent with observations for massive stars of
  any given initial mass.

  \citet{Ptuetal10} also attempted to model the cosmic-ray
  acceleration for four types of SNRs, including three core-collapse types
  and Type-Ia SN. However, the environments these authors assumed were not as detailed as
  ours, and it does not appear that they incorporated the density transitions
  that characterize the circumstellar medium in our
  work. As we have shown, the effects of these transitions make a large
  difference to the particle spectra and the radiation signatures.

We advanced our model of particle acceleration and propagation in
SNRs \citep{Teletal12a, Teletal12b} by adding realistic transport of the
MF. Our calculations suggest that the MF at
the reverse shock is sufficient to accelerate particles to TeV
energies. Although the maximum energy of cosmic-rays at the reverse
shock is lower than at the FS, their intensity is high on account
  of the high ejecta density in core-collapse SNRs. Thus the reverse
shock contributes a significant fraction of the total CR intensity in
SNRs at early times. The RS-accelerated particles soften the total
emission spectra, in agreement with recent data.

Emission spectra from core-collapse SNRs are complex and reflect
  both the distribution of gas or MF and the spectral
  differences between the particle populations coming from the forward
  and reverse shock. High-energy emission from Type-Ic SNR is
  dominated by leptonic processes, particularly so when the SNR is
  older. High-energy radiation from Type-IIP SNRs is very soft and
  strongly dominated by pion-decay emission, except for emission at
  the highest energies from old SNR.

Both types of remnants show, in addition to a shell structure, a
  center-filled morphology of high-energy emission, because the dense
  and weakly magnetized ejecta in the interior permits high-energy
  particles to propagate and radiate there. The shell-type morphology
  of Type-IIP SNR, however, may be difficult to resolve with the
  current generation of gamma-ray telescopes.  We also found that the
  complex radial distribution of MF causes significant
  differences in the morphology of synchrotron and IC emission from
  core-collapse SNRs, and therefore a ``re-mapping'' of the spectra
  and morphology of the two types of leptonic emission is not
  advised.

The high-energy emission from Type-Ic SNRs is roughly an order of
magnitude fainter than that of Type-IIP SNRs at early ages, but
  Type-Ic SNRs are good targets for the current generation of
  telescopes at the age of a thousand years and later, whereas
  Type-IIP SNRs lose brightness very rapidly.

Our work demonstrates that the hydrodynamics of SNRs and their
interaction with the environment play a significant role in shaping
particle and emission spectra as well as intensity maps. The emission
from core-collapse SNRs is significantly different from that of
Type-Ia SNRs that evolve in a uniform environment \citep{Teletal12a}.

\section*{Appendix}

To derive Eq.~\ref{BrBt} we rewrite Eq.~\ref{dotprod} in spherical
coordinates and use the properties of the spherically symmetric flow
profile with velocity, $\mathbf{u}$: $u_r = u_r$, $u_{\theta} = 0$,
$u_{\phi} = 0$ and $\partial u_r/ \partial \theta= 0$, $\partial u_r/
\partial \phi = 0$. The MF, $\mathbf{B}$, has all three
components, $B_r $, $B_{\theta}$, $B_{\phi}$, and $\bigtriangledown
\cdot \mathbf{B} = 0$, therefore

\begin{equation}
\mathbf{u}(\bigtriangledown \cdot \mathbf{B}) = 0
\end{equation}

\begin{equation}
\begin{array}{ll}
\mathbf{B}(\bigtriangledown \cdot \mathbf{u}) = \\
\mathbf{B} \left(
{1 \over r^2}{\partial \left( r^2 u_r \right) \over \partial r} +
{1 \over r\sin\theta}{\partial \over \partial \theta} \left(  u_\theta\sin\theta \right) +  
{1 \over r\sin\theta}{\partial u_\phi \over \partial \phi}
\right) = \\
B_r \left( \frac{2u_r}{r} + \frac{\partial u_r}{\partial r} \right) \boldsymbol{\hat{r}} +
B_{\theta} \left( \frac{2u_r}{r} + \frac{\partial u_r}{\partial r} \right) \boldsymbol{\hat{\theta}}+ 
B_{\phi} \left( \frac{2u_r}{r} + \frac{\partial u_r}{\partial r} \right) \boldsymbol{\hat{\phi}}
\end{array}
\end{equation}

\begin{equation}
\begin{array}{ll}
(\mathbf{B} \cdot \bigtriangledown) \mathbf{u} = \\
\left(B_r \frac{\partial u_r}{\partial r} + 
\frac{B_\theta}{r}\frac{\partial u_r}{\partial \theta} +
\frac{B_\phi}{r\sin(\theta)}\frac{\partial u_r}{\partial \phi} - 
\frac{B_\theta u_\theta + 
B_\phi u_\phi}{r}\right) \boldsymbol{\hat r}  + \\
\left(B_r \frac{\partial u_\theta}{\partial r} + 
\frac{B_\theta}{r}\frac{\partial u_\theta}{\partial \theta} + 
\frac{B_\phi}{r\sin(\theta)}\frac{\partial u_\theta}{\partial \phi} + 
\frac{B_\theta u_r}{r} - 
\frac{B_\phi u_\phi\cot(\theta)}{r}\right) \boldsymbol{\hat\theta} + \\
\left(B_r \frac{\partial u_\phi}{\partial r} + 
\frac{B_\theta}{r}\frac{\partial u_\phi}{\partial \theta} + 
\frac{B_\phi}{r\sin(\theta)}\frac{\partial u_\phi}{\partial \phi} +
\frac{B_\phi u_r}{r} + 
\frac{ B_\phi u_\theta \cot(\theta)}{r}\right)\boldsymbol{\hat\phi} = \\
B_r \frac{\partial u_r}{\partial r} \boldsymbol{\hat{r}} +
B_\theta \frac{u_{r}}{r}  \boldsymbol{\hat{\theta}} + 
B_\phi \frac{u_{r}}{r} \boldsymbol{\hat{\phi}}
\end{array}
\end{equation}

\begin{equation}
\begin{array}{ll}
(\mathbf{u} \cdot \bigtriangledown) \mathbf{B} = \\
\left(u_r \frac{\partial B_r}{\partial r} + 
\frac{u_\theta}{r}\frac{\partial B_r}{\partial \theta} +
\frac{u_\phi}{r\sin(\theta)}\frac{\partial B_r}{\partial \phi} - 
\frac{u_\theta B_\theta + 
u_\phi B_\phi}{r}\right) \boldsymbol{\hat r}  + \\
\left(u_r \frac{\partial B_\theta}{\partial r} + 
\frac{u_\theta}{r}\frac{\partial B_\theta}{\partial \theta} + 
\frac{u_\phi}{r\sin(\theta)}\frac{\partial B_\theta}{\partial \phi} + 
\frac{u_\theta B_r}{r} - 
\frac{u_\phi B_\phi\cot(\theta)}{r}\right) \boldsymbol{\hat\theta} + \\
\left(u_r \frac{\partial B_\phi}{\partial r} + 
\frac{u_\theta}{r}\frac{\partial B_\phi}{\partial \theta} + 
\frac{u_\phi}{r\sin(\theta)}\frac{\partial B_\phi}{\partial \phi} +
\frac{u_\phi B_r}{r} + 
\frac{ u_\phi B_\theta \cot(\theta)}{r}\right)\boldsymbol{\hat\phi} = \\
u_r \frac{\partial B_r}{\partial r} \boldsymbol{\hat{r}} +
u_r \frac{\partial B_{\theta}}{\partial r}  \boldsymbol{\hat{\theta}} + 
u_r \frac{\partial B_{\phi}}{\partial r} \boldsymbol{\hat{\phi}}.
\end{array}
\end{equation}

Combining the expressions, we find for the evolution of the three components of $\mathbf{B}$ 
\begin{equation}
\begin{array}{ll}
\frac{\partial B_r}{\partial t} = 
-2 B_r \frac{u_r}{r} - 
B_r \frac{\partial u_r}{\partial r} + 
B_r \frac{\partial u_r}{\partial r} - 
u_r \frac{\partial B_r}{\partial r} \\
= -
\frac{\partial}{\partial r} \left(B_r u_r\right) -
2 B_r \frac{u_r}{r} +
B_r \frac{\partial u_r}{\partial r} \ ,\\
 \\
\frac{\partial B_{\theta}}{\partial t} =
-2 B_{\theta} \frac{u_r}{r} -
B_{\theta} \frac{\partial u_r}{\partial r} + 
B_{\theta} \frac{u_r}{r} -
u_r \frac{\partial B_{\theta}}{\partial r} = - 
\frac{\partial}{\partial r} \left(B_{\theta} u_r\right) -
B_{\theta} \frac{u_r}{r} \ ,\\
 \\
\frac{\partial B_{\phi}}{\partial t} =
-2 B_{\phi} \frac{u_r}{r} -
B_{\phi} \frac{\partial u_r}{\partial r} + 
B_{\phi} \frac{u_r}{r} -
u_r \frac{\partial B_{\phi}}{\partial r} = - 
\frac{\partial}{\partial r} \left(B_{\phi} u_r\right) -
B_{\phi} \frac{u_r}{r}\ .
\end{array}
\end{equation}

\acknowledgements 
We acknowledge support by the ``Helmholtz Alliance for Astroparticle Physics HAP'' funded by the Initiative and Networking Fund of the Helmholtz Association. VVD's work is supported by NASA Fermi grant NNX12A057G.

\bibliographystyle{aa}
\bibliography{ccsnr}

\end{document}